\documentclass[12pt]{article}
\usepackage[left=2.5cm,top=2.50cm,right=2.5cm,bottom=2.50cm]{geometry}
\usepackage{mathrsfs}
\usepackage{amsmath,amssymb,latexsym,color,cancel,graphicx,colortbl}
\usepackage[english]{babel}
\usepackage[latin1]{inputenc}
\usepackage{ragged2e}
\usepackage{cite}
\begin{document}
\date{}

\title{New Exact Supersymmetric Wave Functions for a Massless Scalar Field and Axion-Dilaton String Cosmology in a FRWL Metric}
\author{R. Cordero$^{a}$  and   R.  D. Mota$^{b}$} \maketitle

\begin{minipage}{0.9\textwidth}
\small $^{a}$ Escuela Superior de F\'{\i}sica y Matem\'aticas,
Instituto Polit\'ecnico Nacional,
Ed. 9, Unidad Profesional Adolfo L\'opez Mateos, Del. Gustavo  A. Madero, C.P. 07738, Ciudad de M\'exico, Mexico.\\

\small $^{b}$ Escuela Superior de Ingenier\'ia Mec\'anica y
El\'ectrica, Unidad Culhuac\'an, Instituto Polit\'ecnico Nacional,  Av. Santa Ana No. 1000, Col. San
Francisco Culhuac\'an, Del. Coyoac\'an,  C.P. 04430,  Ciudad de M\'exico, Mexico.\\
\end{minipage}

\rm
\begin{abstract}
We study the quantum cosmology of the $2D$ Hawking-Page massless scalar-field model and the homogeneous and isotropic dilation-axion string cosmology.
For both models we find new exact solutions for the Wheeler-DeWitt equation and for the supersymmetric wave function equations.
We show that the Wheeler-DeWitt equations for the Hawking-Page model and the axion-dilaton string cosmology are invariant under a one-parameter and two-parameter squeezed transformations, respectively. These results allow to find new solutions for the wave functions of these systems at the quantum cosmology level. Also, we apply the same squeezed transformations to solve the supersymmetric wave equations and find new supersymmetric solutions for the $2D$ Hawking-Page massless scalar-field model and the homogeneous and isotropic dilation-axion fields. Besides, for the former case, with the help of separation of variables we find new supersymmetric wave functions.\\

\end{abstract}

PACS numbers: 98.80.Qc, 11.30.Pb, 02.30.Jr

\section{Introduction}

Quantum cosmology applies the principles of quantum physics to the
Universe as a whole. The wave function of the Universe depending
on the configuration space with its degrees of freedom (superspace) evolves
according to the Wheeler-DeWitt (WDW) equation, which is an
infinite-dimensional partial differential equation. Because of the
serious  difficulties in solving this functional equation, some
solutions can be found in four dimensions in the so called
minisuperspace approximation, where the inhomogeneous modes are
frozen out before quantization. In the beginning of 1980 started an important development of
quantum cosmology when it was introduced the idea that the Universe could be spontaneously
nucleated out of nothing \cite{7,77}, where nothing means the absence of space and
time. Once the nucleation took place the universe follows an inflationary expansion and maintains
its evolution to the present. Nevertheless there are many relevant questions that remain to be
solved like the right boundary conditions for the Wheeler-DeWitt equation. In 4-dimensional quantum
cosmology there is nothing external to the Universe and the boundary conditions cannot be imposed
safely, which it is not the case for quantum mechanics, and the question of the correct boundary condition
remains unsolved.
Several possibilities for such boundary conditions have been presented in Refs. \cite{8,9,10,11,12,LindeT}.
Different physical settings emerge if we choose different boundary conditions, e. g. if the wave function
 is exponentially damped for large three-geometries and is regular when the three-geometry collapses
to zero we have wormholes. Hawking and Page \cite{HAW} considered a
minimally coupled massless scalar field and found its wormhole
solution.

It is well known that symmetries play a central role to understand
many properties of physical systems. They form the basis for
selection rules that forbid the existence of certain states and
processes. Also, from a given solution, symmetries allow us to
obtain new solutions of the field equations. Moreover, the
conserved quantity associated with a given symmetry represent an
integrability condition.

Graham \cite{GRAHAM2} found that there exists an $N=2$ hidden
supersymmetry for the bosonic part $H_0$ of the
supersymmetric Hamiltonian in the
minisuperspace approximation for the Bianchi type IX metric. It has been found that many cosmological models, such as
Bianchi type-II model \cite{mexicanos}, scalar-tensor cosmology
\cite{LIDSEY2} and the two-dimensional dilaton-gravity model
\cite{LIDSEY1}, possess hidden supersymmetry.
The supercharges $Q$ and $\bar Q$  at
quantum level are operators and depend on the Grassmann variables.
The supersymmetric wave function $\Psi_s$ is
annihilated by the supercharges, $Q\Psi_s=0$ and ${\bar
Q}\Psi_s=0$. These supercharges constraints correspond to the
Dirac-type square root of the WDW equation. Therefore,
a set of coupled linear differential equations has to be solved
to find $\Psi_s$. This approach allows to avoid the second-order
WDW equation that needs suitable boundary
conditions \cite{8,9,10,11,12} which is an unsolved problem in 4-dimensional quantum cosmology. This is one of the must important
advantages of using supersymmetry in quantum cosmology.

On the other hand, in the string cosmological scenario, the
dilaton-graviton action is invariant under the scale factor
duality, which is a subgroup of the $O(d,d)$ T-duality group,
being $d$ the spatial number of dimensions. This scale factor duality leads
to a supersymmetric extension of quantum  cosmology
\cite{Lidsey-slac, maha1,maha2}.
The low energy four dimensional
effective theory action of string theory  contains two
massless fields \cite{st}. One of these scalar fields is called
the axion $\chi$ and it comes from the third rank field strength
corresponding to the Kalb-Ramond field, the other one is called
the dilaton $\phi$. The physical consequences of the axion in a
curved spacetime has been investigated in the aim of finding
possible indirect evidences of low energy string theory
\cite{axion1,sen,das,axion2, axion3}. The dilaton is very important in
string theory since it defines the string coupling constant $g_s$
as $e^{\phi/2}$, it determines the Newton constant, the gauge
coupling constants and the Yukawa couplings.

In this paper we focus our attention to find the supersymmetric wave
function  of two very similar systems, at least at the level of the WDW equation.  We study the $2D$
model of the wormhole massless scalar field worked out by Hawking and Page \cite{HAW},
which has been emerged in other cosmological contexts, as two-dimensional
dilaton-gravity models \cite{LIDSEY1}, and in scale factor duality and supersymmetry in scalar-tensor cosmology  \cite{LIDSEY2}.
Besides, we study the FRWL  string cosmology model proposed by Maharana \cite{maha1,maha2}.
For both models we find several closed solutions of the WDW equation and for the supersymmetric wave function.

The paper is organized as follows. In Section 2 we show that de WDW equations of the $2D$ massless scalar field model and the axion-dilaton isotropic and homogeneous string fields are invariant under the so-called squeezed transformation (Lorentz transformation) and  we use this fact to find new solutions for the wave functions for both models. In Section 3 we give the main results of the $N=2$ supersymmetric quantum cosmology. Besides, we obtain new supersymmetric wave function solutions for the $2D$ massless scalar field model and the axion-dilaton string fields. By means of the squeezed transformations we find other new supersymmetric solutions for both models. The method of separation of variables is applied to obtain other set of of SUSY wave functions for the axion-dilaton string cosmology. In Section 4 we report the unnormalized probability density for the supersymmetric quantum cosmology cases.  Finally, we give our concluding remarks.

\section{Quantum cosmology for scalar fields}

\subsection{ $2D$ Massless Scalar-Field Model}
The spacetime corresponding for the models that we study in this paper is described by
the homogeneous and isotropic Friedmann-Robertson-Walker-Lemaitre (FRWL) metric for closed universes ($k=1$)
\begin{equation}
ds^2=-dt^2+a(t)^2\left(\frac{dr^2}{1-r^2}+r^2d\Omega^2\right),
\end{equation}
where  $t$ is the cosmic time.

Although our goal is to study the axion-dilaton string cosmology, we begin with
a Hamiltonian which is very similar to the string cosmology Hamiltonian, at least mathematically.
This $2D$ model is the wormhole massless scalar field studied by Hawking and Page \cite{HAW}.
We notice that this model has emerged in other cosmological contexts, as two-dimensional
dilaton-gravity models \cite{LIDSEY1}, and scale factor duality and supersymmetry in scalar-tensor cosmology  \cite{LIDSEY2}. The Wheeler-DeWitt (WDW) equation
is given by
\begin{equation}
{\mathcal H}_0\psi=\frac{1}{2}\left(\frac{\partial^2}{\partial
y^2}-\frac{\partial^2}{\partial x^2}-(y^2-x^2)\right)\psi=0,
\label{hpham}
\end{equation}
where ${\mathcal H}_0$ is the Hamiltonian and the variables $x$ and $y$ are defined in terms of the massless scalar field $\phi$ and the scale factor $a(t)$ by the relations $x=a\sinh{\phi}$, $y=a\cosh{\phi}$.

Some solutions for this WDW equation are immediately found in terms of the harmonic oscillator Hermite functions, which are given by
\begin{equation}
\psi=H_n(x)H_n(y)e^{-\frac{1}{2}(x^2+y^2)},\label{hermites2d}
\end{equation}
where $n \in \mathbb{N}$. These are the wave functions for the ground state ($E=0$). Since $n$ is an arbitrary integer number,
there exists and infinite degeneracy  to this energy. Also, the parity of the Hermite polynomials $(-1)^n$
implies that $\psi$ is invariant under the duality transformation $x\rightarrow -x$ and $y\rightarrow -y$. \\

If we consider the possibility of matter-energy renormalization \cite{8} by introducing an arbitrary
constant, the WDW equation takes the form
\begin{equation}
\frac{1}{2}\left(-\frac{\partial^2}{\partial
y^2}+\frac{\partial^2}{\partial x^2}+(y^2-x^2)-2E\right)\psi=0.
\label{2dham}
\end{equation}
This equation admits analytical solutions if we consider that the constant $E$ can be quantized. In this case, the solutions are
\begin{equation}
\psi=H_m(x)H_n(y)e^{-\frac{1}{2}(x^2+y^2)},\hspace{3ex} E=n-m.\label{enren1}
\end{equation}

By introducing the light-cone coordinates $x=\frac{1}{\sqrt{2}}(u-v)$ y $y=\frac{1}{\sqrt{2}}(u+v)$, the WDW equation
(\ref{2dham}) can be written as
\begin{equation}
\left(-\frac{\partial^2}{\partial u\partial v}+uv-E\right)\psi=0.
\label{squeezedeq}
\end{equation}
We find that the solutions to this equation are
\begin{equation}
\psi=H_n\left(\frac{u+v}{\sqrt{2}}\right)H_m\left(\frac{u-v}{\sqrt{2}}\right)e^{-\frac{1}{2}(u^2+v^2)},\hspace{3ex}E=n-m.
\end{equation}
It is very important to note that equation (\ref{squeezedeq}) is invariant under the so-called squeezed transformation
\begin{equation}
u' = e^{-\eta}u, \hspace{2cm} v' = e^{\eta}v,
\label{squeezedt}
\end{equation}
and therefore, the squeezed-type functions
\begin{equation}
\psi_\eta=H_n\left(\frac{e^{-\eta}u+e^{\eta}v}{\sqrt{2}}\right)H_m\left(\frac{e^{-\eta}u-e^{\eta}v}{\sqrt{2}}\right)e^{-\frac{1}{2}(e^{-2\eta}u^2+e^{2\eta}v^2)},
\label{solsqueezeduv}
\end{equation}
are eigenfunctions for the WDW equation with energy  $E=n$, being $\eta$ one parameter which takes arbitrary real values $\eta\in \mathbb{R}$. These functions have been emerged in the context of Lorentz boots \cite{kim}. \\

In fact, it is straightforward to show that the squeezed transformation (\ref{squeezedt}) is just the Lorentz transformation
\begin{eqnarray}
x' &=& x\cosh \eta  - y\sinh \eta \nonumber \\
y' &=& - x\sinh \eta  + y\cosh \eta \label{lorentz}
\end{eqnarray}
which leaves invariant the WDW operator
\begin{equation}
\frac{\partial^2}{\partial y^2}-\frac{\partial^2}{\partial
x^2}-(y^2-x^2) = \frac{\partial^2}{\partial {y'}^2}-\frac{\partial^2}{\partial
{x'}^2}-({y'}^2-{x'}^2).
\end{equation}
Thus, because of this, from equation (\ref{enren1}), the function
\begin{equation}
\psi_{\eta}=H_m(x')H_n(y')e^{-\frac{1}{2}(x'^2+y'^2)}.\label{solprimada}
\end{equation}
is a solution of the Wheeler-DeWitt equation (\ref{2dham}), which is the same solution (\ref{solsqueezeduv}), but written in the variables $x'$ and $y'$. Notice that because of the squeezed transformation (\ref {lorentz}), we have label these solution as depending on the free squeezed parameter $\eta$.
Explicitly, the solution (\ref{solprimada} ) is
\begin{equation}
\psi_\eta=H_m(x\cosh \eta  - y\sinh \eta )H_n(- x\sinh \eta  + y\cosh \eta )e^{-\frac{1}{2}(x^2+y^2)\cosh2\eta+2xy\sinh\eta\cosh\eta},
\end{equation}
which is a non-separable solution in the variables $x$ and $y$ with energy $E=n-m.$\\

On the other hand, if we propose $\psi=e^{\frac{1}{2}x^2}g(y)$, the
substitution into the WDW equation leads to
\begin{equation}
\frac{dg(y)}{dy^2}-y^2g(y)-g(y)=0,
\end{equation}
whose solution is given by $g(y)=c_1e^{\frac{1}{2}y^2}+c_2e^{\frac{1}{2}y^2}erf(y)$.
Similarly, if we propose $\psi=e^{-\frac{1}{2}x^2}g(y)$, we find that $g(y)=c_1e^{-\frac{1}{2}y^2}+c_2e^{-\frac{1}{2}y^2}erf(iy)$. Although these wave functions are not normalizable, it is possible to extract some physical information using conditional probability, see for example Ref. \cite{Halliwell,DeLorenci:1997hu}.
\\

Performing the change of variables to the scale factor and the scalar field in the Hamiltonian  (\ref{hpham}),
we obtain
\begin{equation}
{\mathcal H}_0\psi= \left(\frac{1}{a}\frac{\partial}{\partial
a}a\frac{\partial}{\partial
a}-\frac{1}{a^2}\frac{\partial^2}{\partial \phi^2}-a^2\right)\psi =0,
\label{oscilahiper}
\end{equation}
whereas the angular momentum operator is
${\mathcal L}=-i\frac{\partial}{\partial
\phi}=-i\left(y\frac{\partial}{\partial
x}+x\frac{\partial}{\partial y}\right)$.
It is easy to show that  ${\mathcal L}$ is a constant of motion for the Hamiltonian defined in equation
(\ref{hpham}). Thus the function $\psi$ must be of the form $\psi=e^{im\phi}f(a)$.
Substituting this form for $\psi$ into equation (\ref{oscilahiper}), it is found that
$f(a)=J_{\pm \frac{im}{2}}(ia^2/2)$.\\

Since we know that the standard Bessel functions are the Fourier coefficients of
$e^{iz\sin{\phi}}$ \cite{Nikiforov}, we find that the wave packets
\begin{equation}
\psi^{pac}_{1}=e^{-\frac{1}{2}a^2 cosh (2\phi)},\hspace{6ex}\psi^{pac}_{2,3}=e^{\pm \frac{1}{2}i a^2 sinh (2\phi)}.
\end{equation}
are solutions for the $2D$ WDW equation.\\

\subsection{Axion--Dilaton Isotropic and Homogeneous String Cosmology}

We give a summary of the Maharana's papers, Refs. \cite{maha1,maha2},
on the FRWL dilaton-axion string cosmology  which are relevant to our work. In these
references it has been considered the effective action
\begin{equation}
S=\int
d^4x\sqrt{-g}\left(R-\frac{1}{2}\partial_\mu\phi\partial^\mu\phi-\frac{1}{2}
e^{2\phi} \partial_\mu\chi\partial^\mu\chi\right),
\end{equation}
where $R$ is the scalar curvature, $\sqrt{-g}$ is the determinant of
the metric $g_{\mu\nu}$, and $\phi$ and $\chi$ are the dilaton and
axion fields, respectively.  It has been shown that this action is invariant under
the $SU(1,1)$ $S$-duality group. The corresponding WDW equation for the axion--dilaton
string  cosmology is given by
\begin{equation}
H\Psi:=\frac{1}{2} \left(\frac{\partial^2}{\partial a^2}
+\frac{p}{a}\frac{\partial}{\partial a}-a^2+\frac{1}{a^2}\hat
{C}\right)\Psi=0, \label{WDWMAHA}
\end{equation}
where in order to solve the ordering ambiguity between $a$ and
$\partial/\partial a$, it was  adopted the prescription that the
resulting Hamiltonian respect invariance of coordinates in
minisuperspace by setting $p=1$. Since the action $S$ is invariant under
the $SU(1,1)$ transformations, also $H$ is invariant under these
transformations. $\hat C$ is the $SU(1,1)$ Casimir operator, which
expressed in the pseudospherical coordinate system
\begin{equation}
x=a \sinh \alpha \cos \beta,\hspace{5ex}y=a \sinh \alpha \sin
\beta, \hspace{5ex}z=a \cosh \alpha,\label{sphericalparabo}
\end{equation}
is just the Laplace-Beltrami operator given by
\begin{equation}
{\hat C}=-\frac{1}{\sinh{\alpha}}\frac{\partial}{\partial
\alpha}\left( \sinh{\alpha}\frac{\partial}{\partial \alpha}\right)
-\frac{1}{\sinh^2{\alpha}}\frac{\partial ^2}{\partial \beta^2}.
\label{angularmaha}
\end{equation}
The axion and dilaton fields can be written in terms of the
pseudospherical coordinates (\ref{sphericalparabo}) as
\begin{equation}
\chi=\frac{\sinh{\alpha}\cos{\beta}}{\cosh{\alpha}+\sinh{\alpha}\sin{\beta}},
\hspace{8ex}e^{-\phi}=\frac{1}{\cosh{\alpha}+\sinh{\alpha}\sin{\beta}}.
\end{equation}
The explicit solutions for the Wheeler-DeWitt constraint
(\ref{WDWMAHA}) on the pseudosphere were obtained from the
$SU(1,1)$ group theory by identifying that the correct series
involved in quantum cosmology is the continuous one \cite{maha2}.
It was shown that
\begin{equation}
Y^m_{-\frac{1}{2}+i\lambda}(\cosh{\alpha},\beta)=e^{im\beta}P^m_{-\frac{1}{2}+i\lambda}(\cosh{\alpha}),
\label{deg2Y}
\end{equation}
are common eigenfunctions for the non-compact operator ${\hat C}$ with eigenvalue $\lambda^2+\frac{1}{4}$ and the compact generator
$-i\partial_{\beta}$, with eigenvalue $m$.
The functions $P^m_{-\frac{1}{2}+i\lambda}(\cosh{\alpha})$ are the associated
Legendre polynomials (also called toroidal functions), $m$ is an integer or a half integer and $\lambda>0$ and real \cite{maha2}.
Notice that for this case, by varying $\lambda$ and  $m$  there exists and
infinite degeneracy.

Thus, by setting the eigensolutions of the Hamiltonian (\ref{WDWMAHA}), with $p=1$,  as
\begin{equation}
\Psi(a,\alpha,\beta)=\tilde{f}(a)Y^m_{-\frac{1}{2}+i\lambda}(\cosh{\alpha},\beta),
\label{deg2}
\end{equation}
it was found that the scale-factor eigenfunctions $\tilde{f}(a)$ result to be
$\tilde{f}(a)=J_{\pm i\frac{\nu}{2}}(ia^2/2)$, where $\nu^2=\lambda^2+\frac{1}{4}$ \cite{maha1,maha2}.

One of the main goals of the present paper is to find new solutions of the WDW equation and the supersymmetric wave
function for the FRWL axion--dilaton string cosmology. Recognizing that equation (\ref{WDWMAHA}) in the
coordinates (\ref{sphericalparabo}) with factor ordering $p=2$  can
be written as
\begin{equation}
H\Psi=\frac{1}{2}\left( \frac{\partial ^2}{\partial
z^2}-\frac{\partial ^2}{\partial y^2}-\frac{\partial ^2}{\partial
x^2}-(z^2-y^2-x^2)\right)\Psi=0, \label{WDWmahacart}
\end{equation}
turns out to be the key step to obtain new solutions of the WDW equation and the supersymmetric wave function of this model. \\

We could like to report a solution analogous to equation (\ref{hermites2d}) for the wormhole model. However, because we
have an odd number of oscillators, in this case, there is no such  a solution.

We find a solution to the WDW equation as
\begin{equation}
\Psi=e^{\pm \frac{1}{2}(x^2-y^2)}\left(C_1\sqrt{z}I_{\frac{1}{4}}(z^2/2)+C_2\sqrt{z}K_{\frac{1}{4}}(z^2/2)\right).\label{fulana}
\end{equation}
In order to have a convergent wave state for $|x|$, $|y|$, $|z|$ $\rightarrow \infty$, we must set  $y\leq |x|$,  $y\geq -|x|$,
$z \geq 0$. If we set $C_2=0$, this function results to be regular at origin, and the exponential decreases faster than the Bessel function increases. \\

Also, other possible solution is
\begin{equation}
\Psi=e^{\pm izy}\left(C_1\sqrt{x}I_{\frac{1}{4}}(x^2/2)+C_2\sqrt{x}K_{\frac{1}{4}}(x^2/2)\right).\label{sutana}
\end{equation}
This solution is oscillatory in the $y-z$ plane. Thus, although we set $C_2=0$ in $\Psi$ to have a regular
behavior at the origin, it is the growth of the Bessel function the responsible for the divergence of $\Psi$ when $x\rightarrow \infty$.
It is interesting to note that although the wave function is not normalizable there exists the possibility to obtain physical information using conditional probability \cite{Halliwell,DeLorenci:1997hu}. The solutions (\ref{fulana})
and (\ref{sutana}) are solutions to the WDW equation (\ref{WDWmahacart}) which are  additional to those reported in Ref. \cite{NOSOTROS}.

If matter-energy renormalization is allowed, the WDW equation (\ref{WDWmahacart}) takes the form
\begin{equation}\frac{1}{2}\left(-\frac{\partial ^2}{\partial
z^2}+\frac{\partial ^2}{\partial y^2}+\frac{\partial ^2}{\partial
x^2}+(z^2-y^2-x^2)-2E\right)\Psi=0. \label{WDWstringME}
\end{equation}
Because of our experience gained in the study of the $2D$ model,  we immediately find that the solutions of this WDW equation are
\begin{equation}
\Psi=H_{m}(z)H_n(y)H_p(x)e^{-\frac{1}{2}(z^2+y^2+x^2)},\hspace{5ex}E=m-n-p-\frac{1}{2}.
\end{equation}

We notice, for example, that the Lorentz transformation obtained from a rotation by an angle $\omega\in [0,2\pi)$ on the $x-y$ plane and a boost along the $x-$axis ($\eta\in \mathbb{R}$)
\begin{eqnarray}
x' &=& x\cosh \eta \cos \omega  + y\cosh \eta \sin \omega  - z\sinh \eta, \nonumber \\
y' &=& - x\sin \omega  + y\cos \omega, \nonumber \\
z' &=& -x\sinh \eta \cos\omega - y\sinh\eta \sin\omega + z\cosh\eta, \label{lorentz2}
\end{eqnarray}
leaves invariant the WDW operator
\begin{eqnarray}
&&-\frac{\partial^2}{\partial z^2}+\frac{\partial^2}{\partial y^2}+\frac{\partial^2}{\partial
x^2}+(z^2 -y^2-x^2) = \nonumber \\
&&-\frac{\partial^2}{\partial z'^2}+\frac{\partial^2}{\partial {y'}^2}-\frac{\partial^2}{\partial
{x'}^2}+(z'^2 -{y'}^2-{x'}^2). \label{WDW3D}
\end{eqnarray}
Thus,
\begin{equation}
\Psi_{\eta\,\omega}=H_{m}(z')H_n(y')H_p(x')e^{-\frac{1}{2}(z'^2+y'^2+x'^2)},\hspace{5ex}E=m-n-p-\frac{1}{2}.
\end{equation}
is a solution. Because of the squeezed transformation (\ref{lorentz2}), we have labeled this solution with the two free parameters $\omega$ and $\eta$. It is possible to perform an additional boost along the $y$ - axis but the explicit expressions for the solutions are cumbersome.

Other solutions are obtained by transforming our Hamiltonian (\ref{WDWmahacart}) to the pseudospherical coordinates (\ref{sphericalparabo}) with $p=2$ in (\ref{WDWMAHA}). We obtain
\begin{equation}
\left(\frac{\partial^2}{\partial a^2}
+\frac{2}{a}\frac{\partial}{\partial a}-a^2+\frac{1}{a^2}\hat
{C}\right)\Psi=0, \label{WDWMAHA2}
\end{equation}
with $\hat{C}$ the Laplace-Beltrami operator (\ref{angularmaha}). If we set $\Psi$ as
\begin{equation}
\Psi(a,\alpha,\beta)=\tilde{\tilde{f}}(a)Y^m_{-\frac{1}{2}+i\lambda}(\cosh{\alpha},\beta),
\label{deg3}
\end{equation}
then, the eigenvalues of the Laplace-Beltrami operator are given by $\lambda^2+\frac{1}{4}$. Thus, we find that the scale-factor eigensolutions are
$\tilde{\tilde{f}}(a)=\sqrt{a}J_{\pm i\lambda}(\frac{1}{2}i a^2)$.

\section{$N=2$ Supersymmetric Approach to Quantum Cosmology}

In the minisuperspace approximation the classical Hamiltonian constraint takes the form
\begin{equation}
2H_0=G^{\mu \nu}p_\mu p_\nu+W(q)=0,
\end{equation}
where the metric $G_{\mu \nu}$ has the signature $(-,+,+,+,...)$, and
$q^\mu$ are the coordinates, $\mu=0,1,...,D$. The $p_\mu$ are the momenta conjugate to
these variables, and $W$ represents the superpotential.

If exists a function $I(q)$  which satisfies
the Hamilton-Jacobi equation
\begin{equation}
W=G^{\mu \nu}\frac{\partial I}{\partial q^\mu}\frac{\partial
I}{\partial q^\nu}, \label{super}
\end{equation}
then, the Hamiltonian ${H}_0$ represents the bosonic
component of a supersymmetric Hamiltonian  \cite{GRAHAM2, LIDSEY2, LIDSEY1, GRAHAM, CLAUDSON, VargasMoniz:2010zz, VargasMoniz:2010zz2}.

At quantum level it is introduced the fermionic degrees of freedom obeying the spinor algebra
\begin{equation}
[\phi^\mu,\phi^\nu]_+=[\bar{\phi}^\mu,\bar{\phi}^\nu]_+=0,
\hspace{6ex}[\phi^\mu,\bar{\phi}^\nu]_+=G^{\mu \nu},
\end{equation}
which allows to write the non-Hermitian supercharges
\begin{equation}
Q=\phi^\mu\left(p_\mu+i\frac{\partial I}{\partial q^\mu}\right),
\hspace{6ex}\bar{Q}=\bar{\phi}^\mu \left(p_\mu-i\frac{\partial
I}{\partial q^\mu}\right),
\end{equation}
satisfying
\begin{equation}
[H_0,Q]_-=[H_0,\bar{Q}]_-=0,\hspace{3ex}\hbox{and }\hspace{3ex}Q^2=\bar{Q}^2=0.\label{susy1}
\end{equation}
Thus, the supercharges $Q$ and $\bar{Q}$ are
conserved.

By choosing the representation
$\bar{\phi}^\mu=\theta^\mu$ and $\phi^\mu=G^{\mu
\nu}\frac{\partial}{\partial \theta^\mu}$ for the fermionic
degrees of freedom, being $\theta^\mu$ Grassmann variables, and the
boson degrees of freedom as $p_\mu=-i\frac{\partial}{\partial
q^\mu}$, the quantized superspace Hamiltonian is given by
\begin{equation}
H=\frac{1}{2}\left[Q,\bar{Q}\right]_+\label{hsusy}={\mathcal H}_0+\frac{1}{2}\frac{\partial^2}{\partial q^\mu
\partial q^\nu}\left[\bar{\phi}^\mu,\phi^\nu \right]_-.
\end{equation}
It can be shown that the last term vanishes
in the classical limit  \cite{GRAHAM2}. Equations (\ref{susy1}) and
(\ref{hsusy}) represent the algebra for the $N=2$ supersymmetry
\cite{LIDSEY2}.

Thus, the supersymmetric WDW constraint means
$H\Psi_s=0$, where $\Psi_s$ is the supersymmetric wave function.
Because of equation (\ref{susy1}), the wave function $\Psi_s$ is
annihilated by the supercharge operators
\begin{equation}
Q\Psi_s=\bar{Q}\Psi_s=0. \label{cargasaniquilantes}
\end{equation}
These constraints represent the so-called square root of the
WDW equation. The supersymmetric wave function was found to have the general
form \cite{GRAHAM2}
\begin{equation}
\Psi_s={\cal A}_++{\cal B}_\nu \theta^\nu+\frac{1}{2}\epsilon_{\nu
\mu \lambda}{\cal C}^\lambda \theta^\nu \theta^\mu + {\cal
A}_-\theta^0\theta^1\theta^2,\label{susygraham}
\end{equation}
where the eight functions ${\cal A}_+$, ${\cal B}_\nu$,  ${\cal
C}_\nu$  and ${\cal A}_-$ depend on the coordinates $x,y,z$, and are given by
\begin{equation}
{\cal A}_\pm=d_\pm e^{\mp I(q)},\hspace{3ex}{\cal
B}_\nu=\frac{\partial f_+(q)}{\partial
q^\nu}e^{-I(q)},\hspace{3ex}{\cal C}^\nu=G^{\nu \mu}
\frac{\partial f_-(q)}{\partial q^\mu}e^{+I(q)},
\label{gradientes}
\end{equation}
where $q^\nu=(z,y,x)$, $G^{\mu \nu}=\hbox{diag}(-1,1,1)$,
$d_{\pm}$ are constants and  $f_{\pm}(q)$ are functions which satisfy
the equations
\begin{equation}
G^{\mu\nu}\left(\frac{\partial}{\partial q^\nu}\mp 2\frac{\partial
I }{\partial q^\nu}\right)\frac{\partial f_\pm (q)}{\partial q^\mu}=0.
\label{supercero}
\end{equation}
The functions $A_+$  and $A_-$ are the empty and filled fermion sectors of the Hilbert space.

\subsection{ $2D$ Massless Scalar-Field Model}

We are going to construct the supersymmetric wave function for the massless scalar field model.
Since for this case $\mu,\nu=0,1$ , $\epsilon_{\mu\nu\lambda}=0$ and in the last term of equation (\ref{susygraham})
the product of the Grassmann variables  runs from 0 to 1.  Thus equation (\ref{susygraham}) reduces to
\begin{equation}
\Psi_s= A_+ +B_0\theta^0+B_1\theta^1+A_-\theta^0\theta^1,\label{susy2d}
\end{equation}
where $q^\mu=(y,x)$ and $G^{\mu\nu}=\hbox{diag}(-1,1)$.

The superpotential is found by solving the Einstein-Hamilton-Jacobi  (\ref{super}) and (\ref{hpham}).
We find
\begin{equation}
I=\frac{1}{2}\left(y^2+x^2\right). \label{super2d}
\end{equation}
We notice that this function is manifestly invariant under the
duality transformation $x\rightarrow -x$ and $y\rightarrow -y$. \\

To find $f_\pm(q)$, we must to solve equation (\ref{supercero}). To this end we use the anzatz,
$f_\pm=W_\pm e^{\pm I}$ introduced in the context of  supersymmetric Bianchi type-II
model \cite{mexicanos}. Thus, equation (\ref{supercero}) leads to
\begin{equation}
\left(\frac{\partial^2}{\partial y^2}-\frac{\partial^2}{\partial
x^2}-(y^2-x^2)\right)W_\pm=0.\label{wmasmen}
\end{equation}
This equation admits the family of solutions \cite{HAW,LIDSEY2}
\begin{equation}
W_\pm=H_n(y)H_n(x)e^{-I}.
\end{equation}
Since in $2D$ the SUSY wave function is given in terms of $B_0$ and $B_1$, we use only $W_+$.

In what follows we need the following properties of the Hermite functions
\begin{eqnarray}
&&\left(\frac{d}{dx_i}+x_i\right)u_n(x_i)=2nu_{n-1}(x_i),\label{H1}\\
&&\left(-\frac{d}{dx_i}+x_i\right)u_n(x_i)=u_{n+1}(x_i), \label{H2}\\
&&H_{n+1}(x_i)=2x_iH_n(x_i)-2nH_{n-1}(x_i),\label{H3}
\end{eqnarray}
with  $u_n(x_i)=H_n(x_i)e^{-\frac{1}{2}x_i^2}$, $x_i=x,y,z$,  \\

By using (\ref{H1}) we can show that the supersymmetric wave function coefficients $B_0$ and $B_1$ are given by
\begin{equation}
B_0=2nH_{n-1}(y)H_n(x)e^{-I},\hspace{6ex}
B_1=2nH_{n}(y)H_{n-1}(x)e^{-I},
\end{equation}
and thus,
\begin{equation}
\psi_{s}=e^{-I}+2n\left[H_{n-1}(y)H_n(x)\theta^0
+H_{n}(y)H_{n-1}(x)\theta^1\right]e^{-I}+e^I\theta^0\theta^1,  \label{2dsusy}
\end{equation}
where we have used second equality of equation (\ref{gradientes}), and $A_\pm =e^{\mp I}$.
$A_\pm$ can be interpreted as the lowest-order WKB approximations of the bosonic Hamiltonian (\ref{hpham})
of the  WDW equation \cite{LIDSEY1}.
We notice that the SUSY wave state (\ref{2dsusy}) has been obtained  in Ref. \cite{LIDSEY2,LIDSEY1} by
solving the equations (\ref{potential01})-(\ref{potencial2}) below. We have obtained them with the ansatz
$f_\pm=W_\pm e^{\pm I}$. The same ansatz permit us to find a SUSY wave function for isotropic and homogeneous string cosmology. \\

Additionally, in what follows, we find a new supersymmetric wave function solution for the wormhole model.
The annihilation of the supersymmetric state (\ref{susy2d}) by the
supercharge operators (\ref{cargasaniquilantes}) translates
into the set of coupled, first-order partial differential
equations
\begin{eqnarray}
&&\left(\frac{\partial}{\partial y}+\frac{\partial I}{\partial y}\right)A_+=0,\hspace{4ex}\left(\frac{\partial}{\partial x}+\frac{\partial I}{\partial x}\right)A_+=0,\label{potential01}\\
&&\left(\frac{\partial}{\partial y}-\frac{\partial I}{\partial y}\right)A_-=0,\hspace{4ex}\left(\frac{\partial}{\partial x}-\frac{\partial I}{\partial x}\right)A_-=0,\label{potential02}\\
&&\left(\frac{\partial}{\partial y}+\frac{\partial I}{\partial y}\right)B_1-\left(\frac{\partial}{\partial x}+\frac{\partial I}{\partial x}\right)B_0=0,\label{potencial1}\\
&&\left(\frac{\partial}{\partial y}-\frac{\partial I}{\partial y}\right)B_0-\left(\frac{\partial}{\partial x}-\frac{\partial I}{\partial x}\right)B_1=0\label{potencial2}.
\end{eqnarray}
The solutions for $A_\pm$ are immediately found to be $A_+=e^{-I}$ and $A_-=e^I$. Introducing new functions defined by $A=B_1 -B_0$, $B=B_0+B_1$ and the new variables
$\alpha=y+x$ and $\beta=y-x$, equations
(\ref{potencial1}) and (\ref{potencial2}) can be rewritten as
\begin{equation}
\frac{\partial B}{\partial \beta}+\frac{\partial I}{\partial \alpha}A=0,\hspace{10ex}\frac{\partial A}{\partial \alpha}+\frac{\partial I}{\partial \beta}B=0.\label{par}
\end{equation}
The uncoupled differential equation for $B$ results to be
\begin{equation}
\left(\frac{\partial^2 }{\partial \alpha \partial \beta}- \frac{1}{\alpha}\frac{\partial }{\partial \beta}-\frac{\alpha\beta}{4}\right)B=0.
\label{soloriginal}
\end{equation}
Since  $I=\frac{\alpha^2+\beta^2}{4}$,  solving this equation by separation of variables, we find
\begin{equation}
B=c_2\alpha e^{\frac{\alpha^2}{8c_1}+\frac{c_1\beta^2}{2}},
\end{equation}
being $c_1$ and $c_2$ arbitrary constants of integration. By using the first of equations (\ref{par}) we find
\begin{equation}
A=-2\beta c_1 c_2e^{\frac{\alpha^2}{8c_1}+\frac{c_1\beta^2}{2}}.
\end{equation}
Using these results and the redefinitions $c_1 \rightarrow \frac{c_1}{2}$ and $\frac{c_2}{2}\rightarrow c_2$, we find
\begin{eqnarray}
B_0=c_2 \left(  \left( 1+c_1 \right) y+ \left( 1-c_1 \right) x \right)  e^{\frac{(y+x)^2}{4c_1}+\frac{c_1(y-x)^2}{4}},\label{b0}\\
B_1=c_2 \left(  \left( 1+c_1 \right) x+ \left( 1-c_1 \right) y \right)  e^{\frac{(y+x)^2}{4c_1}+\frac{c_1(y-x)^2}{4}}\label{b1}.
\end{eqnarray}
Thus, with the above results, we have found the new supersymmetric wave function for the wormhole model
\begin{equation}
\psi=e^{-I}+B_0\theta^0+B_1\theta^1+e^I\theta^0\theta^1.\label{new}
\end{equation}
In order to have solutions which are bounded for $|x|$,$|y|$ $\rightarrow \infty$ we must set $c_1<0$.

It is very important to note that the operator in equation (\ref{soloriginal}) and the equations (\ref{par}) rewritten in the form
\begin{equation}
A=-\frac{2}{\alpha}\frac{\partial B}{\partial \beta},\hspace{6ex} B=-\frac{2}{\beta}\frac{\partial A}{\partial \alpha}
\end{equation}
are invariant under the squeezed transformation $\alpha'=e^{-\eta}\alpha$ and $\beta'=e^\beta\eta$. Since $\alpha=y+x$ and $\beta=y-x$, this transformation is equivalent to
\begin{eqnarray}
x'=x\cosh\eta-y\sinh\eta=\frac{1}{2}\left(\alpha e^{-\eta}-\beta e^{\eta}\right),\label{squeezed1}\\
y'=-x\sinh\eta+y\cosh\eta=\frac{1}{2}\left(\alpha e^{-\eta}+\beta e^{\eta}\right)\label{squeezed2}.
\end{eqnarray}
Thus, any solutions for $A$ and $B$ in prime coordinates lead to the solutions $B_0(x',y')=\frac{1}{2}(B(x',y')-A(x',y'))$ and $B_1(x',y')=\frac{1}{2}(A(x',y')+B(x',y'))$
for equations (\ref{potencial1}) and (\ref{potencial2}) in the  $x'$- and $y'$-coordinates. Because of the invariance we pointed out, by using the transformations (\ref{squeezed1}) and (\ref{squeezed2}) we could obtain $B_0(x',y')$ and $B_1(x',y')$ which also must be solutions for equations (\ref{potencial1}) and (\ref{potencial2}) in $x$- and $y$-coordinates. Based on this fact, if we set
\begin{eqnarray}
&&B_0 = 2n H_{n-1}(y')H_n(x')e^{-I'}, \label{newsqueezed0}\\
&&B_1 = 2n H_{n-1}(x')H_n(y')e^{-I'}, \label{newsqueezed1}\\
&&I'=\frac{1}{2}\left((x')^2+(y')^2\right)\nonumber
\end{eqnarray}
then, by written them in the $x$- and $y$-coordinates (using (\ref{squeezed1}) and (\ref{squeezed2})) together with
\begin{equation}
I'=\frac{1}{2}(x^2+y^2)\cosh2\eta-2xy\sinh\eta\cosh\eta,\label{superprima}
\end{equation}
it can be shown that they satisfy equations (\ref{potencial1}) and (\ref{potencial2}).
Also, by setting $n=1$ and $c_1=-e^{2\eta}$, we show that they reduce in full agreement to $B_0$ and $B_1$ given by equations (\ref{b0}) and (\ref{b1}), respectively. In this way we have shown that the supersymmwtric solution (\ref{new}) is a very particular SUSY solution corresponding to $B_0$ and $B_1$ of equations (\ref{newsqueezed0}) and (\ref{newsqueezed1}) in the $x-y$ coordinates.

Now, if we use the squeezed transformation directly to the solution (\ref{b0}) and (\ref{b1}) in order to obtain a new solution, we just obtain the same solution with a new constant $c_1 \rightarrow e^{-2\eta}c_1$.

On the other hand, we notice that the differential operator of equation (\ref{wmasmen}) is invariant under the squeezed transformation (\ref{lorentz}) (or that given by equations (\ref{squeezed1}) and (\ref{squeezed2})). Thus because of the discussion of Sec. 2.1 its solution in the $x'$- and $y'$-coordinates are
\[
 W_\pm=H_n(y')H_n(x')e^{-I'}.
\]
By using the second expression of equation (\ref{gradientes}), with $q^\mu=(y,x)$ and $G^{\mu \nu}=\hbox{diag}(-1,1)$, we obtain
\begin{eqnarray}
 B_0 &=& c[((1 -\cosh 2\eta )y + x\sinh 2\eta ) H_n(x')H_n(y') + 2n \cosh \eta H_{n-1}(y')H_n(x') \nonumber \\
 &-& 2n \sinh \eta H_{n}(y')H_{n-1}(x')]e^{-I'}, \nonumber
\end{eqnarray}
\begin{eqnarray}
 B_1 &=& c[((1 -\cosh 2\eta )x + y\sinh 2\eta ) H_n(x')H_n(y') + 2n \cosh \eta H_{n-1}(x')H_n(y') \nonumber \\
 &-& 2n \sinh \eta H_{n}(x')H_{n-1}(y')]e^{-I'}. \nonumber
\end{eqnarray}
which can be written using the recurrence relation (\ref{H3}) for Hermite polynomials  as
\begin{equation}
B_0 = c( \sinh\eta H_n(y')H_{n+1}(x') + 2n \cosh \eta H_n(x')H_{n-1}(y'))e^{-I'}, \label{newsqueezedw0}
\end{equation}
\begin{equation}
B_1 = c( \sinh\eta H_n(x')H_{n+1}(y') + 2n \cosh \eta H_n(y')H_{n-1}(x'))e^{-I'}. \label{newsqueezedw1}
\end{equation}
In these expressions $x'$, $y'$ and $I'$ are given by equations (\ref{squeezed1}), (\ref{squeezed2}), and (\ref{superprima}), respectively.
It is important to note that equations (\ref{newsqueezed0})-(\ref{newsqueezed1}) and equations (\ref{newsqueezedw0})-(\ref{newsqueezedw1}) are each other's independent solutions. Also, we point out that by setting $n=0$ and $c_1=-e^{2\eta}$, equations (\ref{newsqueezedw0})-(\ref{newsqueezedw1}) reduce to (\ref{b0}) and (\ref{b1}) as a particular case.

We can observe from the last result that the Hermite polynomials $H_n(y')H_{n+1}(x')$ and $H_n(x')H_{n+1}(y')$ could be considered as a basis for the solutions of the wave functions. The more general supersymmetric wave functions can be written as
\begin{eqnarray}
B_0 = \sum_{n=1} ^{\infty} c_n H_n(x')H_{n-1}(y')e^{-I'} \label{newsqueezedt0} \\
B_1 = \sum_{n=1} ^{\infty} c_n H_n(y')H_{n-1}(x')e^{-I'} \label{newsqueezedt1}.
\end{eqnarray}
In section 5, related to the density of probability,  we plot some typical unnormalized supersymmetric  wave functions for (\ref{newsqueezedw0}) and (\ref{newsqueezedw1}). The plots of the solutions (\ref{newsqueezed0}) and (\ref{newsqueezed1}) are very similar to those reported in Section 5 of the paper.

\subsection{Axion--Dilaton Isotropic and Homogeneous String Cosmology}

We are going to construct the supersymmetric ground state wave function for this model. By solving  the corresponding Hamilton-Jacobi to this equation, we get the superpotential
\begin{equation}
I=\frac{1}{2}\left(z^2+y^2+x^2\right),
\end{equation}
and thus, the bosonic and fermionic components of the ground state
$A_+$ and $A_-$ are obtained in a straightforward way by means of the first relation in equation (\ref{gradientes}). Also, we notice that in this case the superpotential is manifestly invariant
under the duality transformation $x\rightarrow -x$, $y\rightarrow
-y$ and $z\rightarrow -z$.\\

On the other hand, the ansatz that we have used to find the supersymmetric wave function of the $2D$ wormhole
model, $f_{\pm}=W_{\pm} e^{\pm I}$,  permit us to reduce equation (\ref{supercero}) to
\begin{equation}
\left(\frac{\partial ^2}{\partial z^2}-\frac{\partial ^2}{\partial
y^2}-\frac{\partial ^2}{\partial
x^2}-(z^2-y^2-x^2)\mp1\right)W_\pm=0, \label{SUSYPOT}
\end{equation}
which is the analogous of the WDW equation obtained for this model in Section 2.2 with energy $E=\pm 1$. We find that the solutions to these equations are
\begin{equation}
W_\pm=H_{2n+\frac{1}{2}\mp\frac{1}{2}}(z)H_n(y)H_n(x)e^{-I}.\label{w3d}
\end{equation}
The ansatz $f_\pm=W_\pm e^{\pm I}$, ($I=\frac{1}{2}(x^2+y^2+z^2)$ for the FRWL string cosmology and $I=\frac{1}{2}(x^2+y^2)$ for the $2D$ massless scalar field model),
allows to write the coefficients for the supersymmetric ground state, equations  (\ref{gradientes}) as follows
\begin{eqnarray}
B_\nu=\left(\frac{\partial}{\partial q^\nu}+\frac{\partial I}{\partial q^\nu}\right)W_+\\
C^\nu=G^{\mu \nu}\left(\frac{\partial}{\partial q^\mu}-\frac{\partial I}{\partial q^\mu}\right)W_-
\end{eqnarray}
This means that the SUSY  ground state coefficients can be obtained by applying the
creation-annihilation operators to the potential-type functions
$W_+$ and $W_-$.\\
If we use the Hermite functions recursion relations (\ref{H1}) and (\ref{H3}), equations (\ref{gradientes}) allow us to
find $B_\nu$ and $C^\nu$. Thus, we obtain
\begin{eqnarray}
&&A_+=e^{-I}\label{super1}\\
&&B_0=4nH_{2n-1}(z)H_n(y)H_n(x)e^{-I},\\
&&B_1=2nH_{2n}(z)H_{n-1}(y)H_n(x)e^{-I},\\
&&B_2=2nH_{2n}(z)H_n(y)H_{n-1}(x)e^{-I},\\
&&C^0=H_{2n+2}(z) H_n(y)H_n(x)e^{-I},\\
&&C^1=-H_{2n+1}(z)H_{n+1}(y)H_n(x)e^{-I},\\
&&C^2=-H_{2n+1}(z)H_n(y)H_{n+1}(x)e^{-I}\\
&&A_-=e^I.\label{super2}
\end{eqnarray}
These are the coefficients which fully determine the supersymmetric wave function $\Psi_s$ of equation (\ref{susygraham})
for the FRWL string cosmology model.

An alternative set of equations  for the supersymmetric ground state are obtained by the
annihilation of the supersymmetric state (\ref{susygraham}) by the
supercharge operators (\ref{cargasaniquilantes}). This leads to the set of coupled, first-order partial differential
equations
\begin{eqnarray}
&&\left(\frac{\partial}{\partial z}+\frac{\partial I}{\partial z}\right)A_+=0,\hspace{4ex}\left(\frac{\partial}{\partial y}+\frac{\partial I}{\partial y}\right)A_+=0,\hspace{4ex}\left(\frac{\partial}{\partial x}+\frac{\partial I}{\partial x}\right)A_+=0,\label{potential001}\\
&&\left(\frac{\partial}{\partial z}-\frac{\partial I}{\partial z}\right)A_-=0,\hspace{4ex}\left(\frac{\partial}{\partial y}-\frac{\partial I}{\partial y}\right)A_-=0,\hspace{4ex}\left(\frac{\partial}{\partial x}-\frac{\partial I}{\partial x}\right)A_-=0\label{potential002}\\
&&\left(\frac{\partial}{\partial z}+\frac{\partial I}{\partial z}\right)B_1-\left(\frac{\partial}{\partial y}+\frac{\partial I}{\partial y}\right)B_0=0,\label{potencial11}\\
&&\left(\frac{\partial}{\partial z}+\frac{\partial I}{\partial z}\right)B_2-\left(\frac{\partial}{\partial x}+\frac{\partial I}{\partial x}\right)B_0=0\label{potencial22},\\
&&\left(\frac{\partial}{\partial y}+\frac{\partial I}{\partial y}\right)B_2-\left(\frac{\partial}{\partial x}+\frac{\partial I}{\partial x}\right)B_1=0\label{potencial23},\\
&&\left(\frac{\partial}{\partial z}-\frac{\partial I}{\partial z}\right)B_0-\left(\frac{\partial}{\partial y}-\frac{\partial I}{\partial y}\right)B_1-\left(\frac{\partial}{\partial x}-\frac{\partial I}{\partial x}\right)B_2=0,\label{potencial24}\\
&&\left(\frac{\partial}{\partial z}-\frac{\partial I}{\partial z}\right)C^1+\left(\frac{\partial}{\partial y}-\frac{\partial I}{\partial y}\right)C^0=0,\label{potencial25}\\
&&\left(\frac{\partial}{\partial z}-\frac{\partial I}{\partial z}\right)C^2+\left(\frac{\partial}{\partial x}-\frac{\partial I}{\partial x}\right)C^0=0\label{potencial26},\\
&&\left(\frac{\partial}{\partial y}-\frac{\partial I}{\partial y}\right)C^2-\left(\frac{\partial}{\partial x}-\frac{\partial I}{\partial x}\right)C^1=0\label{potencial27},\\
&&\left(\frac{\partial}{\partial z}+\frac{\partial I}{\partial z}\right)C^0+\left(\frac{\partial}{\partial y}+\frac{\partial I}{\partial y}\right)C^1+\left(\frac{\partial}{\partial x}+\frac{\partial I}{\partial x}\right)C^2=0.\label{potencial28}
\end{eqnarray}
This set of equations is the analogous one to that for the wormhole model (\ref{potential01})-(\ref{potencial2}) introduced by Lidsey \cite{LIDSEY2,LIDSEY1} for the $2D$ wormhole model. By using the recursion relations (\ref{H1}) and (\ref{H2}) it is immediate to show that our functions (\ref{super1})-(\ref{super2}) satisfy this set of differential equations.

To obtain another set of solutions to equations  (\ref{potencial11})-(\ref{potencial28}) we propose the separation of variables $B_0=B_0(z)B_0(y)B_0(x)$, $C^0=C^0(z)C^0(y)C^0(x)$ and similar expression for $B_1$, $B_2$, $C^1$ and $C^2$.
After some calculations we find that
\begin{eqnarray}
&&B_0=c H_{n}(z)H_{n+1-m}(y)H_m(x)e^{-I},\label{rub1}\\
&&B_1=c \frac{n+1-m}{n+1}H_{n+1}(z)H_{n-m}(y)H_m(x)e^{-I},\label{rub2}\\
&&B_2=c \frac{m}{n+1}H_{n+1}(z)H_{n+1-m}(y)H_{m-1}(x)e^{-I},\label{rub3}\\
&&C^0=d H_{n+3}(z)H_{n+1-m}(y)H_m(x)e^{-I},\label{rub4}\\
&&C^1=-d H_{n+2}(z)H_{n+2-m}(y)H_m(x)e^{-I},\label{rub5}\\
&&C^2=-d H_{n+2}(z)H_{n+1-m}(y)H_{m+1}(x)e^{-I},\label{rub6}
\end{eqnarray}
being $c$  and $d$ arbitrary constants. We emphasize that the set of equations (\ref{potencial11})-(\ref{potencial28}) for the supersymmetric ground state
of the FRWL string cosmology accept more general solutions that those which are obtained from equations (\ref{susygraham})- (\ref{supercero}) as it is shown in equations (\ref{rub1})-(\ref{rub6}).
Notice that if we set $n=2n'-1$,  $m=n'$,  $c=4n'$  and  $d=1$ our SUSY ground state coefficients reduce in full agreement to those reported in equations
(\ref{super1})-(\ref{super2}).

Since equation (\ref{SUSYPOT}) is invariant under the Lorentz transformation (\ref{lorentz2}), we proceed as in Section 3.1 for the two-dimensional case to obtain other set of supersymmetric solutions. $W_\pm$ in prime coordinates is $W_\pm=cH_{2n+\frac{1}{2}\mp\frac{1}{2}}(z')H_n(y')H_n(x')e^{-I'}$. Therefore, $f_\pm=W_\pm e^{\pm I}$, leads to
\begin{eqnarray}
&&f_+=cH_{2n}(z')H_n(y')H_n(x')e^{I-I'},\\
&&f_-=cH_{2n+1}(z')H_n(y')H_n(x')e^{-(I'+I)}.
\end{eqnarray}
Since $B_0=\frac{\partial f_+}{\partial z}e^{-I}$, $B_1=\frac{\partial f_+}{\partial y}e^{-I}$ and $B_2=\frac{\partial f_+}{\partial x}e^{-I}$, we obtain
\begin{eqnarray}
B_0 &=& c(4n \cosh \eta H_{2n-1}(z')H_n(y')H_n(x') -2n\sinh\eta H_{2n}(z')H_n(y')H_{n-1}(x') \nonumber \\
 &+& 2\sinh\eta(-z\sinh\eta  + x\cos\omega\cosh\eta  + y\sin\omega\cosh\eta )H_{2n}(z')H_n(y')H_n(x'))e^{-I'}, \label{3d1}\\
B_1 &=& c(-4n \sinh \eta \sin\omega H_{2n-1}(z')H_n(y')H_n(x') +2n\cos\omega H_{2n}(z')H_n(x')H_{n-1}(y') \nonumber \\
  &+& 2\sinh\eta\sin\omega(z\cosh\eta  - x\cos\omega\sinh\eta  - y\sin\omega\sinh\eta )H_{2n}(z')H_n(y')H_n(x') \nonumber \\
  &+& 2n\cosh\eta \sin\omega H_{2n}(z')H_n(y')H_{n-1}(x'))e^{-I'}, \\
B_2 &=& c(-4n \sinh \eta \cos\omega H_{2n-1}(z')H_n(y')H_n(x')-2n\sin\omega H_{2n}(z')H_n(x')H_{n-1}(y') \nonumber \\
  &+& 2\sinh\eta\cos\omega(z\cosh\eta  - x\cos\omega\sinh\eta  - y\sin\omega\sinh\eta )H_{2n}(z')H_n(y')H_n(x') \nonumber \\
  &+& 2n\cosh\eta \cos\omega H_{2n}(z')H_n(y')H_{n-1}(x'))e^{-I'}.
\end{eqnarray}
Similarly, since $C^0=-\frac{\partial f_-}{\partial z}e^{I}$, $C^1=\frac{\partial f_-}{\partial y}e^{I}$ and $C^2=\frac{\partial f_-}{\partial x}e^{I}$, we obtain
\begin{eqnarray}
C^0 &=& -c(2(2n+1)\cosh \eta H_{2n}(z')H_n(y')H_n(x') -2n\sinh\eta H_{2n+1}(z')H_n(y')H_{n-1}(x') \nonumber \\
 &-& 2\cosh\eta(z\cosh\eta  - x\cos\omega\sinh\eta  - y\sin\omega\sinh\eta )H_{2n+1}(z')H_n(y')H_n(x'))e^{-I'}, \\
C^1 &=& c(-2(2n+1) \sinh \eta \sin\omega H_{2n}(z')H_n(y')H_n(x') +2n\cos\omega H_{2n+1}(z')H_{n-1}(y')H_n(x') \nonumber \\
  &+& 2\sinh\eta\sin\omega(z\cosh\eta  - x\cos\omega\sinh\eta ) - 2y(\sin^2 \omega\sinh ^2\eta+1))H_{2n+1}(z')H_n(y')H_n(x') \nonumber \\
  &+& 2n\cosh\eta \sin\omega H_{2n+1}(z')H_n(y')H_{n-1}(x'))e^{-I'}, \\
C^2 &=& c(-2(2n+1) \sinh \eta \cos \omega H_{2n}(z')H_n(y')H_n(x')-2n\sin \omega H_{2n+1}(z')H_{n-1}(y')H_n(x') \nonumber \\
  &+& (2\sinh\eta\cos\omega(z\cosh\eta  - y\sin\omega\sinh\eta ) -2x(\cos ^2\omega\sinh ^2\eta +1))H_{2n+1}(z')H_n(y')H_n(x') \nonumber \\
  &+& 2n\cosh\eta \cos\omega H_{2n+1}(z')H_n(y')H_{n-1}(x'))e^{-I'}\label{3d2},
\end{eqnarray}
In equations (\ref{3d1})-(\ref{3d2}) $I'$ is given by
\begin{eqnarray}
I'=\frac{1}{2}((2\cosh^2\eta-1)z^2+(2\sinh^2\eta\cos^2\omega+1)x^2+(2\sinh^2\eta\sin^2\omega+1)y^2)\nonumber\\
+2xy\sinh^2\eta\cos\omega\sin\omega-2xz\sinh\eta\cosh\eta\cos\omega-2yz\sinh\eta\cosh\eta\sin\omega.
\end{eqnarray}
The above results is a set of non-separable solutions in the variables $x, y, z$ for the supersymmetric wave function. These solutions are the generalization to the three-dimensional case of the solutions given in (\ref{newsqueezedw0}) and (\ref{newsqueezedw1}). Surprisingly, the corresponding squeezed version solutions of the equations  (\ref{rub1})-(\ref{rub6}) are not solutions for the supersymmetric wave functions which could be expected to be solutions in view of the results obtained in (\ref{newsqueezed0}) and (\ref{newsqueezed1}) for the massless scalar field case.

On the other hand, if we propose the anzatz $B_0=e^{-I}\partial_z\phi$,  $B_1=e^{-I}\partial_y\phi$ and $B_2=e^{-I}\partial_x\phi$, with
$\phi=X(x)+Y(y)+Z(z)$, we  find the non-normalizable solutions
\begin{eqnarray}
&&B_0={\tilde a}\alpha e^{-\frac{1}{2}(x^2+y^2-z^2)} erf(z),\label{rub12}\\
&&B_1={\tilde b}\beta e^{-\frac{1}{2}(x^2+z^2-y^2)} erf(y),\label{rub22}\\
&&B_2={\tilde c}\gamma e^{-\frac{1}{2}(y^2+z^2-x^2)} erf(x)\label{rub32}.
\end{eqnarray}
with $\alpha=\beta+\gamma$, ${\tilde a}$, ${\tilde b}$,  ${\tilde c}$ constants, and $erf(x)$ the standard error function. Similarly, the anzatz  $C^0=e^{I}\partial_z\psi$,  $C^1=-e^{I}\partial_y\psi$ and $C^2=-e^{I}\partial_x\psi$, with
$\psi={\tilde X}(x)+{\tilde Y}(y)+{\tilde Z}(z)$, we  find
\begin{eqnarray}
&&C^0={\tilde d}\delta e^{\frac{1}{2}(x^2+y^2-z^2)} fer(z),\label{rub42}\\
&&C^1=-{\tilde e}\epsilon e^{\frac{1}{2}(x^2+z^2-y^2)} fer(y),\label{rub52}\\
&&C^2=-{\tilde f}\omega e^{\frac{1}{2}(y^2+z^2-x^2)} fer(x).\label{rub62}
\end{eqnarray}
with $\delta=\epsilon+\omega$, ${\tilde d}$, ${\tilde e}$,  ${\tilde f}$ constants, and $fer(x)\equiv \frac{2}{\sqrt{\pi}}\int_0^x{e^{t^2}dt}$.
Although the above wave functions are not normalizable, it is possible to get physical information by means of the conditional probability \cite{Halliwell,DeLorenci:1997hu}.

\section{The unnormalized probability density $|\Psi|^2$}

The unnormalized probability density of the supersymmetric wave function $|\Psi|^2$ is obtained by integrating over
the Grassmann variables. It was shown that  it is given by {\cite{mexicanos}
\begin{equation}
|\Psi|^2={\cal A}_+^* {\cal A}_++{\cal B}_0^*{\cal B}_0+{\cal B}_1^*{\cal B}_1+{\cal B}_2^*{\cal B}_2+{\cal C}^{0*}{\cal C}^0+{\cal C}^{1*}{\cal C}^1+{\cal C}^{2*}{\cal C}^2 + {\cal A}_-^*{\cal A}_-.
\end{equation}

\begin{figure}
\begin{center}
\begin{tabular}{lll}
\includegraphics[scale=0.4]{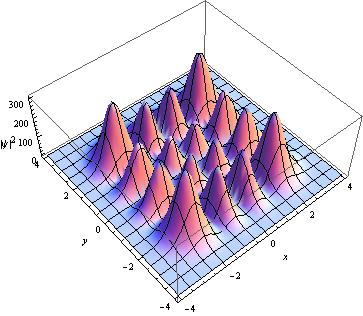}
\includegraphics[scale=0.4]{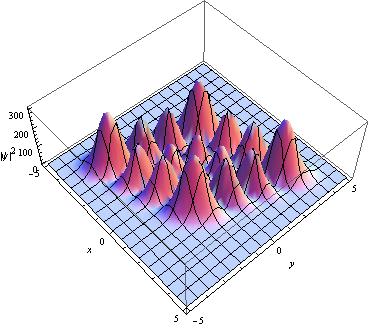}
\includegraphics[scale=0.4]{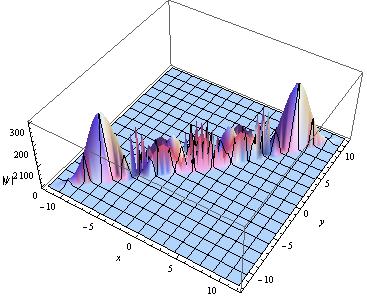}
\end{tabular}
\caption{Plot of the unnormalized $|\Psi|^2$ for the two dimensional model, $B_0$ and $B_1$ given by (\ref{newsqueezed0}) and (\ref{newsqueezed1}). For $n=3$, we have set $\eta=0$, $\eta=0.3$ and $\eta=1.3$}.
\end{center}
\end{figure}

\begin{figure}
\begin{center}
\begin{tabular}{lll}
\includegraphics[scale=0.4]{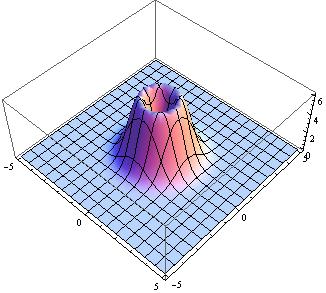}
\includegraphics[scale=0.4]{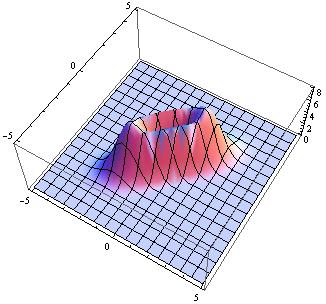}
\includegraphics[scale=0.4]{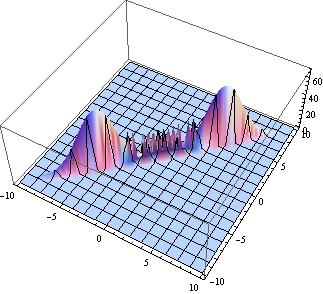}
\end{tabular}
\caption{Plot of the unnormalized $|\Psi|^2$ 2D supersymmetric solutions given  by Eqs. (\ref{newsqueezedw0}) and (\ref{newsqueezedw1}) with $n=0$. We have set $\eta=0$, $\eta=0.3$ and $\eta=1.3$}
\end{center}
\end{figure}

\begin{figure}
\begin{center}
\begin{tabular}{lll}
\includegraphics[scale=0.4]{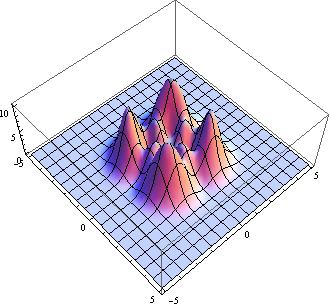}
\includegraphics[scale=0.4]{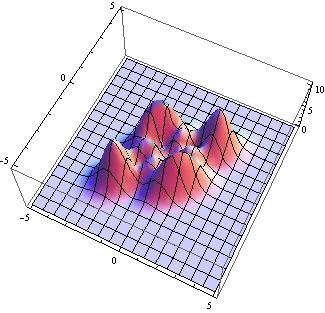}
\includegraphics[scale=0.4]{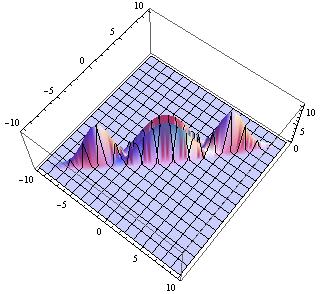}
\end{tabular}
\caption{Plot of the unnormalized $|\Psi|^2$ 2D supersymmetric solutions given  by Eqs. (\ref{newsqueezedw0}) and (\ref{newsqueezedw1}) with $n=1$. We have set $\eta=0$, $\eta=0.3$ and $\eta=1.3$}
\end{center}
\end{figure}

\begin{figure}
\begin{center}
\begin{tabular}{lll}
\includegraphics[scale=0.4]{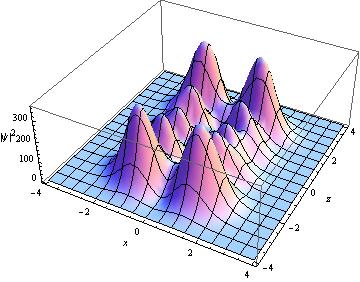}
\includegraphics[scale=0.4]{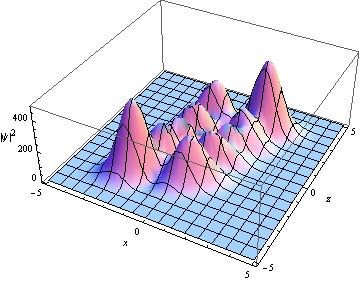}
\includegraphics[scale=0.4]{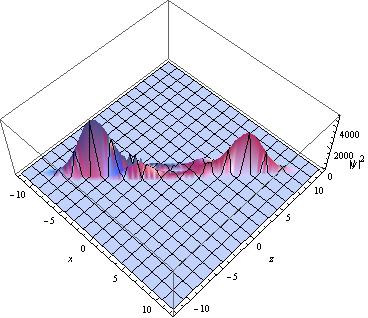}
\end{tabular}
\caption{Unnormalized $|\Psi|^2$ 3D supersymmetric solutions given  by Eqs. (\ref{3d1})-(\ref{3d2}). For $n=1$, $y=0.5$, $\omega=0$, we have set $\eta=0$, $\eta=0.3$ and $\eta=1.3$}
\end{center}
\end{figure}

\begin{figure}
\begin{center}
\begin{tabular}{lll}
\includegraphics[scale=0.4]{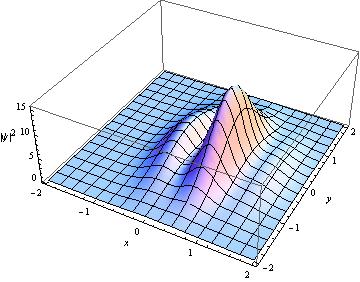}
\includegraphics[scale=0.4]{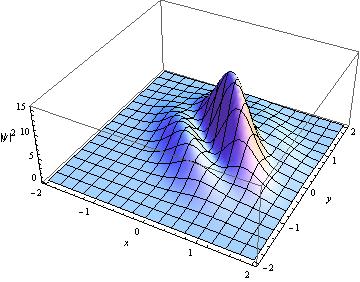}
\includegraphics[scale=0.4]{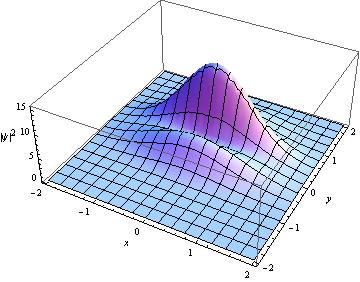}
\end{tabular}
\caption{Unnormalized $|\Psi|^2$ 3D supersymmetric solutions given  by Eqs. (\ref{3d1})-(\ref{3d2}). For $n=0$, $z=0.5$, $\eta=1$, we have set $\omega=0$, $\omega=\pi/4$ and $\omega=\pi/2$}
\end{center}
\end{figure}

In all the plots we report in this paper for the Hawking-Page scalar model, by demanding that $|\Psi|^2$ be finite when $|x|$,$|y|$ $\rightarrow \infty$ we have neglected the term $e^I$ in the SUSY wave function. In Fig. 1, we have plotted the unnormalized probability density for the supersymmetric wave function of the two-dimensional Hawking-Page scalar model. For this case, $B_0$ and $B_1$ are given by (\ref{newsqueezed0}) and (\ref{newsqueezed1}). For $n=3$, we have set $\eta=0$, $\eta=0.3$ and $\eta=1.3$. In Figs. 2 and 3, we have plotted the unnormalized $|\Psi|^2$ 2D supersymmetric solutions given  by equations (\ref{newsqueezedw0}) and (\ref{newsqueezedw1}) with $\eta=0$, $\eta=0.3$ and $\eta=1.3$ for $n=0$ and $n=1$, respectively. Also, we have plotted the unnormalized probability density $|\Psi|^2$ for the supersymmetric axion-dilaton string cosmology solutions given  by Eqs. (\ref{3d1})-(\ref{3d2}).
We observe that the coefficients remains finite as  $|x|$, $|y|$, $|z|$ $\rightarrow \infty$, except  the last term ($e^I$) of the SUSY wave function of equation (\ref{susygraham}). Fig. 4, shows the probability density for $n=1$, $y=0.5$ and $\omega=0$ fixed, and varying $\eta$ to take the values 0, 0.3 and 1.3, respectively.
From the plots for the probability density we deduce that the squeezed supersymmetric solutions for the dilaton-axion cosmology possesses an analogous behavior as that for the massless scalar field. In fact, for both models the probability density is located mainly along the line $x=y$ and this effect is enhance when $n$ increases. When $\eta$ takes negative values the probability density is located mainly along the line $x=-y$. In fact, these results are to be expected from the functional dependence of the exponential in terms of the light-cone coordinates $\alpha$ and $\beta$. In fig. 5, we have fixed the values $n=0$, $z=0.5$, and $\eta=1$ and we have choose $\omega$ to take the values $0$, $\pi/4$ and $\pi/2$. This shows that in effect, the rotation along the $z-$axis in space leads to a rotation of the probability density.

\section{Concluding Remarks}

In this paper we have studied the quantum cosmology of the $2D$ Hawking-Page massless scalar field model and the homogeneous and isotropic dilation-axion string cosmology and we have obtained new solutions to their corresponding Wheeler-DeWitt equation.

Besides, we have found new supersymmetric wave functions for the Hawking-Page scalar field model and the FRWL string cosmology by means of the so-called squeezed transformation and by separation of variables. For the Hawking-Page scalar field model we were able to decoupled the differential equations that fulfill the functions $B_0$ and $B_1$ by means of the introduction of a new set of functions and the light cone coordinates. We get new solutions for $B_0$ and $B_1$ which are not written as a product of functions of the original variables. However this solution corresponds to a particular case of the solutions obtained by the squeezed transformations. In the case of the homogeneous and isotropic dilation-axion string cosmology, the solutions obtained by applying directly the method of separation of variables to the set of equations that satisfies the functions $Bs$ and $Cs$ contain as a particular case the solutions found by using the procedure of Graham \cite{GRAHAM2}. It is remarkable that, as far as we know, this is the first time that it was possible to obtain this more general solution for the axion-dilaton case. 

On the other hand, one of the more important features of the squeezed solutions is that they are not written as a product of functions of the original variables. Our new supersymmetric probability density possesses a rich variety of possibilities in its behavior which is very different from the well known standard solution. From the probability density plots we conclude that the supersymmetric solutions for the dilaton-axion cosmology and the massless scalar field possesses an analogous behavior. In fact, for both models the probability density is located mainly along the line $x=y$ and this effect is enhance when $n$ increases. When $\eta$ takes negative values the probability density is located mainly along the line $x=-y$. Also, from the plots we deduce that the rotation along the $z-$axis in space leads to a rotation of the probability density.

In quantum cosmology the wave function of the Universe is constructed from suitable combinations of the solutions of the Wheeler-DeWitt equation depending on the prescriptions for boundary conditions that one adopt. 
Also, we point out that we can introduce a more general squeezed transformation depending on three-free parameters, obtained from a rotation around the $z-$axis and independent boosts along the $x-$ and $y-$axis, which leaves invariant the WDW operator (\ref{WDW3D}). This would allow to obtain more general and complete solutions than those we have reported. Because of what we have seen in our analysis above, we can conjecture that for this case, a more rich behavior of the probability density would exist due to the squeezing effects.  Finally, we emphasize that our procedure can be applied immediately to find the Wheeler-DeWitt solutions and the supersymmetric wave functions for multidimensional quantum wormholes studied in Refs. \cite{HAW1,ZHUK}. These issues will be studied in a future work.

\section*{Acknowledgments}

This work was partially supported by SNI-M\'exico, COFAA-IPN, EDI-IPN, SIP-IPN 20180741 and 20195330 number projects.

\end{document}